\documentclass[aps, preprint]{revtex4}

\usepackage{graphicx}
\usepackage{color}

\definecolor{purple}{rgb}{1,0,1}
\definecolor{brown}{rgb}{0.45,0,0}
\definecolor{lightblue}{cmyk}{0.75,0,0,0}
\definecolor{lightred}{cmyk}{0,0.1,0,0}
\definecolor{lightyellow}{cmyk}{0,0,0.5,0}

\def\beq{\begin{equation}}
\def\eeq{\end{equation}}
\def\bea{\begin{eqnarray}}
\def\eea{\end{eqnarray}}

\def\nn{\nonumber}
\def\sss{\scriptscriptstyle}

\def\barp{{\raise.35ex\hbox
{${\sss (}$}}---{\raise.35ex\hbox{${\sss )}$}}}
\def\barpd{{\raise.35ex\hbox
{${\sss (}$}}--{\raise.35ex\hbox{${\sss )}$}}}
\def\dbarp{\hbox{$D^{**0}$\kern-1.9em\raise1.5ex\hbox{\barpd}}}
\def\bdbarp{\hbox{$B_d$\kern-1.4em\raise1.4ex\hbox{\barp}}}
\def\bsbarp{\hbox{$B_s$\kern-1.4em\raise1.4ex\hbox{\barp}}}

\def\roughly#1{\mathrel{\raise.3ex\hbox
{$#1$\kern-.75em\lower1ex\hbox{$\sim$}}}}

\def\adir00{{a_{\sss dir}^{00}}}

\def\B00{B^{00}}
\def\Bp0{B^{+0}}

\def\dsp{\displaystyle}

\begin{document}

\preprint{IMSc-2004/05/21}

\title{\Large \bf \boldmath Determining $\gamma$ using $B\to D^{**}K$}

\author{Nita Sinha}\email{nita@imsc.res.in}
\affiliation{The Institute of Mathematical Sciences, C.I.T Campus, Taramani, 
Chennai 600113, India.} 
\date{\today}

\begin{abstract} 
{We propose the use of $B^\pm\to \dbarp~~~K^\pm$
decay modes for a theoretically clean determination of the weak phase
$\gamma$. The self tagging decays of the neutral $D^{**}$ mesons,
makes a measurement of the $b\to u\bar{c}s$ amplitude feasible. This
overcomes the problem with the Gronau-London-Wyler proposal. Even an
upper limit on the ${\cal B}(B^-\to\bar{D}^{**0}K^-)$ will place an
assumption free, lower bound on $\gamma$.}
\end{abstract}

\maketitle

\newpage

It is hoped that theoretically clean, precise measurements of all the
angles of the unitarity triangle will provide us with a testing ground
for the Standard Model (SM) parameterization of $CP$
violation~\cite{Harrison:1998yr}. A time dependent $CP$ asymmetry in
the golden mode $B^0\to J/\psi K_s$ has been successfully used to
measure $\sin(2\beta)$~\cite{Browder:2003ii}.  Clean methods to
determine {\em all} the $CP$ violating phases from a variety of final
states are crucial in a search for physics beyond the SM. A clean
extraction of the weak phase $\gamma$ has been an experimental
challenge.  While first estimates of the angle $\gamma$ are provided
by methods based on approximations like SU(3)~\cite{Gronau:2003br},
precise measurements have to be free from such assumptions. A
promising theoretically clean method was proposed by Gronau, London
and Wyler (GLW)~\cite{GLW}.  The method relies on the interference of
the $b\to c\bar{u}s$ and $b\to u\bar{c}s$ tree amplitudes. The former
appears in the decay of $B^-\to D^0K^-$, while the latter in $B^-\to
\bar{D}^0K^-$. If one observes the decay $B^-\to D_{CP}K^-$, where
$D_{CP}$ is a $CP$ eigenstate into which the neutral $D$ decays, then
the interference achieved allows determination of $\gamma$. The
technique requires a measurement of the branching ratios: ${\cal
B}(B^-\to D^0K^-)$, ${\cal B}(B^-\to \bar{D}^0K^-)$, $ {\cal B}(B^-\to
D_{CP}K^-)$ and $ {\cal B}(B^+\to D_{CP}K^+)$.  However, the
measurement of ${\cal B}(B^-\to \bar{D}^0K^-)$ poses an experimental
problem~\cite{Atwood:1996ci}. The branching ratio ${\cal B}(B^-\to
\bar{D}^0K^-)$ measured by reconstructing the $\bar{D}^0$ in a
hadronic mode is contaminated by ${\cal B}(B^-\to D^0K^-)$, where the
$D^0$ decays via a doubly-Cabibbo suppressed decay mode. The
$\bar{D}^0$ cannot be tagged through a semileptonic mode either, due
to the background from direct decays of the $B^-$.

A resolution to this problem was provided by Atwood, Dunietz and
Soni~\cite{Atwood:1996ci}, who considered decays into two final states
of the neutral $D$, with at least one of the final states being not a
$CP$ eigenstate. The ${\cal B}(B^-\to \bar{D}^0K^-)$ is treated as a
parameter that can be solved. The method has the advantage that the
$CP$ asymmetry for the non-$CP$ eigenstate is larger, since the final
state is chosen to be a doubly Cabibbo suppressed mode of
$D^0$. However, a precise measurement of $\gamma$, from this method
requires at least one precise determination of a doubly Cabibbo
suppressed branching ratio of $D^0$.  Another method using the
vector-vector $D^*K^*$ final states was
proposed~\cite{Sinha:1997zu}. An angular analysis results in a large
number of observables that allow determination of $\gamma$ as well as
other parameters, including the doubly Cabibbo suppressed branching
ratio.  Several other variations of the $DK$ method have also been
presented~\cite{dk}. Recently, alternate methods using either
multibody final states or multilple final states of the $D^0$ have been
investigated~\cite{recent}. While the sensitivity of most $DK$ methods
is expected to be similar, larger number of methods are useful for
increasing statistics and providing consistency checks~\cite{soffer}.

In this brief report, we propose the use of $B^\pm \to \dbarp~~~
K^\pm$ decays, as these will allow an implementation of the GLW
method.  Since the excited $D$-meson states, $D^{**0}$
($\bar{D}^{**0}$), decay into a charged $D^+\pi^-/D^{*+}\pi^-$
($D^-\pi^+/D^{*-}\pi^+$), the branching ratios for both $B^- \to
D^{**0} K^\pm$ as well as $B^- \to \bar{D}^{**0} K^\pm$ can be
measured. Hence, the difficulty of the GLW method is overcome. Note
that the vector $D^{*0}$ meson cannot decay into charged $D\pi$ modes
and hence, using $D^*K^\pm$ final states does not resolve the problem.

Various $D^{**0}$ mesons have been observed by many
collaborations~\cite{expt}. The Belle collaboration has recently
measured the product of branching fractions: ${\cal B}(B^-\to
D^{**0}\pi^-) \times {\cal B}(D^{**0}\to (D/D^*)^+\pi^-) \approx
O(10^{-4})$, for four of the D-meson excited states~\cite{belle}. With
larger number of $B-\bar{B}$ pairs expected, it should soon be
possible to measure the branching ratios for the $D^{**0}K^-$ final
state as well. The high luminosity at the super-B factories and other
planned B-physics experiments, should enable a measurement of the
branching fractions for ${\cal B}(B^-\to \bar{D}^{**0}K^-) \times
{\cal B}(\bar{D}^{**0}\to (D/D^*)^-\pi^+)$, allowing $\gamma$ to be
cleanly extracted. Even if only an upper limit on ${\cal B}(B^-\to
\bar{D}^{**0}K^-)$ is available, it will still allow us to obtain a
bound on $|\sin\gamma|$, free of any theoretical assumptions.

The decay amplitudes for $B^-$ may be defined as:
\begin{eqnarray}
  \label{amps}
{\cal A}(B^-\to {D}^{**0}K^-\to D^+\pi^-K^-) &\equiv& a_c e^{i\delta_c}\nn\\
{\cal A}(B^-\to \bar {D}^{**0}K^-\to D^-\pi^+K^-) &\equiv& a_u e^{i\delta_u} e^{-i\gamma},
\end{eqnarray}
where, $a_c$, $a_u$ are the decay amplitudes involving the $b\to c\bar{u}s$ and
$b\to u\bar{c}s$ transitions and $\delta_c$, $\delta_u$ are the
corresponding strong phases. In the Wolfenstein
parameterization~\cite{Wolfenstein:1983yz}, while the amplitude for
$B^-\to D^{**0}K^-$ has no weak phase, that for $B^-\to \bar
{D}^{**0}K^-$ has the weak phase $\gamma$. Interference of these
amplitudes is achieved by looking at the decays into $CP$ eigenstates of the neutral $D$, $B^-\to D_{CP}^\pm \pi^0 K^-$,
\begin{equation}
  \label{amps2}
{\cal A}(B^-\to [D^{**0}K^-\pm\bar {D}^{**0}K^-]\to D_{CP}^\pm\pi^0K^-)=\frac{r}{\sqrt{2}}[a_c e^{i\delta_c}\pm
a_u e^{i\delta_u} e^{-i\gamma}],
\end{equation}
where, $r$ is the ratio of the amplitudes of $D^0$ to a $CP$ eigenstate to
that of the $D^+$ to a Cabibbo-allowed mode (or any mode in which it
is reconstructed) and the $CP$ even (odd) eigenstates of $D^0$ are
defined as: $D_{CP}^\pm=\dsp\frac{1}{\sqrt{2}}[D^0\pm\bar{D}^0]$.
Using these and the
amplitudes for the $CP$ conjugate modes, we have:
\begin{eqnarray}
  \label{obs}
{\cal B}_{sum} &\equiv& {\cal B}(B^-\to \big[D_{CP}^\pm\pi^0\big]_{D^{**}}K^-)+{\cal B}(B^+\to
\big[D_{CP}^\pm\pi^0\big]_{D^{**}}K^+)\nn\\
 &=&r^2(a_c^2+a_u^2\pm 2a_c a_u
\cos\delta \cos\gamma)
\end{eqnarray}
and the $CP$ asymmetry,
\begin{eqnarray}
  \label{obs2}
A_{CP}&\equiv& \frac{{\cal B}(B^-\to
\big[D_{CP}^\pm\pi^0\big]_{D^{**}}K^-)-{\cal B}(B^+\to
\big[D_{CP}^\pm\pi^0\big]_{D^{**}}K^+)}{{\cal B}(B^-\to
\big[D_{CP}^\pm\pi^0\big]_{D^{**}}K^-)+{\cal B}(B^+\to
\big[D_{CP}^\pm\pi^0\big]_{D^{**}}K^+)} \nn\\
 &=& \frac{\mp 2a_u
a_c\sin\delta\sin\gamma}{a_c^2+a_u^2\pm 2a_c a_u \cos\delta \cos\gamma}
\end{eqnarray}
where, $\delta=\delta_c-\delta_u$.
 
The measured values for ${\cal B}(B^-\to {D}^{**0}K^-)$ and ${\cal
B}(B^-\to \bar {D}^{**0}K^-)$ determine $a_c$ and $a_u$
respectively. While the corresponding amplitude $a_c$ could be
determined in the original GLW method, a measurement to determine
their corresponding $a_u$, was not feasible.  Our choice of final
states with excited neutral $D$ mesons that decay into self tagging
charged $D\pi$ modes, has made the determination of $a_u$ possible.
Knowing $a_c$ and $a_u$ and the two observables in Eqs.(\ref{obs} and
\ref{obs2}), the phases, $\delta$ and $\gamma$ can be
determined. $|\sin\gamma|$ is determined up to a two-fold ambiguity
from the relation:
\begin{equation}
 \label{singamma}
\sin^2\gamma=\frac{4r^4a_u^2a_c^2+A_{CP}^2{\cal B}_{sum}^2-X^2\mp\sqrt{(4r^4a_u^2a_c^2+A_{CP}^2{\cal B}_{sum}^2-X^2)^2-16r^4a_u^2a_c^2A_{CP}^2{\cal B}_{sum}^2}}{8r^4a_u^2a_c^2}
\end{equation}
where, $X \equiv {\cal B}_{sum}-r^2a_c^2-r^2a_u^2$.

Note that the technique presented is also applicable to the decay of
the neutral $B$ mesons, $\bar{B}^0\to D^{**0}\bar{K}^{*0} (B^0\to
\bar{D}^{**0}K^{*0})$. The decay of $K^*$ into self tagging modes
render the neutral $B$ decay modes to be treated exactly like the
charged $B$ decays. While the branching ratios in case of the neutral
$B$ decays are expected to be smaller, the $CP$ asymmetry will be
larger as both the $b\to c$ as well as $b\to u$ contributions will be
colour suppressed and therefore comparable. Only inclusive data in the
$\bar{B}^0\to D^{**0}\bar{K}^{*0} (B^0\to \bar{D}^{**0}K^{*0})$ modes
will be required; an angular analysis to obtain individual partial
wave amplitudes need not be performed. In the decays of neutral
$B$ mesons, the usefulness of $D^{**}$ states, had been pointed out in
Ref.~\cite{Kayser:1999bu}, where a method to extract
$(2\beta+\gamma)$, using a time dependent study was presented.

Several of the $D^{**0}$ modes have been observed and the product of
branching ratios for production of $D^{**0}\pi$ and the decay of
$D^{**0}\to D\pi (D^*\pi)$ have been measured.  In principle, any of
the $D^{**0}$ modes could be used.  Since the $CP$ asymmetry will be
larger for the mode with larger strong phase difference $\delta$, this
variety of $D^{**0}$ states will allow one to choose the state with
the largest $CP$ asymmetry, large branching ratio and high
reconstruction efficiency. For our numerical estimations below, as an
example, we choose the $D_2^*$ resonance, since this has a narrow
width and decays to a $D^+\pi^-$. The axial vector mesons, $D_1^0$,
$D_1^{'0}$ are allowed to decay only into $D^{*+}\pi^-$.  Hence, the
reconstruction efficiency for the axial vector mesons might be
expected to be lower than that for the tensor or the scalar. Moreover
there is also the problem of mixing in case of the axial vector
mesons. The scalar $D_0^{*0}$ is rather broad. A possible complication
is that both the $D_0^{*0}$ and the $D_2^{*0}$ have overlapping Breit
Wigner shapes.  However, due to the narrow width of $D_2^{*0}$, it may
be possible to select the region of interference and extract the
region corresponding entirely to $D_0^{*0}$~\cite{Abi}.

In the following, we show, that even if the ${\cal B}(B^-\to \bar
{D}^{**0}K^-)$ is not exactly measured but only an upper limit is
available, one can already start putting bounds on $|\sin\gamma|$. For
this numerical analysis we use the following~\cite{numbers}:
\begin{eqnarray}
a_c^2 &=& \lambda^2 {\cal B}(B^-\to D_2^{*0}\pi^-) {\cal B}(D_2^{*0}\to
      D^+\pi^-) {\cal B}(D^+\to K^-\pi^+\pi^+)= 2.69\times10^{-6}\nn\\
r^2 &=& \frac{{\cal B}(D^0\to K_S\pi^0)}{{\cal B}(D^+\to K^-\pi^+\pi^+)}= 0.125\nn\\
{\cal B}_{sum} &=& 3.45\times10^{-7}.
\label{nvalues}
\end{eqnarray} 
Naively, we expect, 
$\dsp\frac{a_u^2}{a_c^2}
=\Big[\frac{|V_{ub}V_{cs}|}{|V_{cb}V_{us}|}\frac{a_2}{a_1}\Big]^2=0.030$,
where, to obtain the colour suppression factor, we use $\Big(\dsp\frac{a_2}{a_1}\Big)^2 =
2~\dsp\frac{B^0\to\bar{D}^0\pi^0}{B^0\to D^-\pi^+}= 0.194$.
\begin{figure}[htb]
\begin{center}
\includegraphics*[scale=1.2]{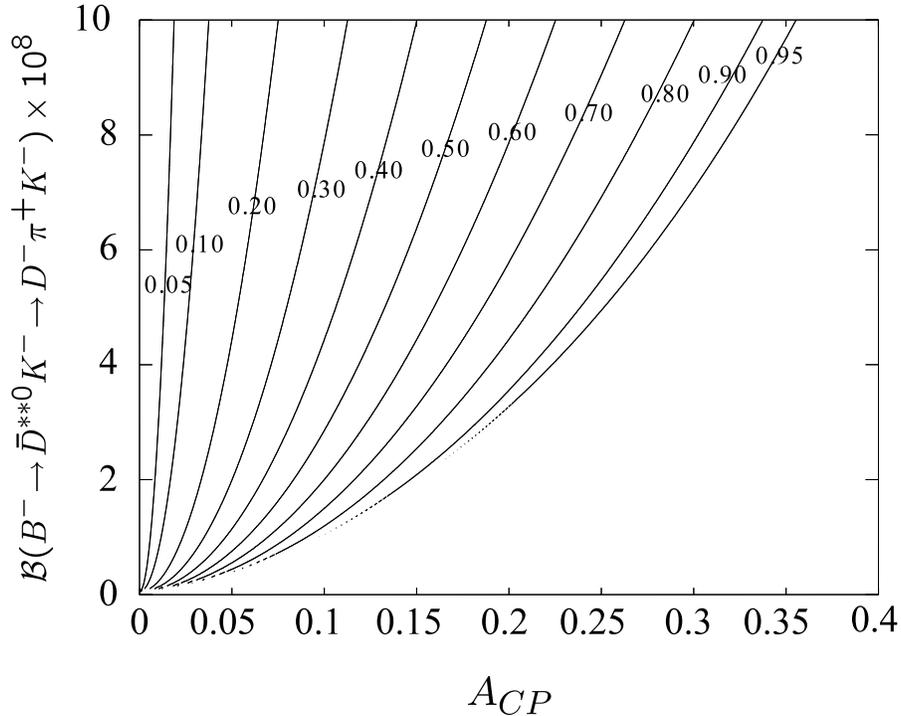}
\caption{$|\sin\gamma|$ contours corresponding to various values of
the $CP$ asymmetry, $A_{CP}$ and the branching ratio, ${\cal B}(B^-\to
\bar {D}^{**0}K^-\to D^-\pi^+K^-)$.}
\label{Fig1}
\end{center}
\end{figure}

Using the numerical values of the parameters given by
Eq.~(\ref{nvalues}), an extremization of the expression for
$\sin^2\gamma$ given in Eq.(\ref{singamma}) results in bounds on
$|\sin\gamma|$ for limiting values of $a_u^2$. We find that while the
first solution (with the -ve sign for the discriminant) gives a lower
bound on $|\sin\gamma|$ for $a_u^2\lesssim 5.4\times 10^{-6}$, the second
solution (with the +ve sign for the discriminant) provides an upper
bound on $|\sin\gamma|$ for $a_u^2\lesssim 7.0\times 10^{-8}$. The first
solution vanishes for $A_{CP}=0$, on the other hand, the second is non-vanishing
for zero $CP$ asymmetry.  The $|\sin\gamma|$ contours corresponding to
the first solution are shown in Fig.~\ref{Fig1}. From the figure,
lower bounds on $|\sin\gamma|$, for a measured upper limit on $a_u^2$
and non-vanishing $CP$ asymmetry can be easily estimated. For example,
if we assume that the measured $a_u^2$ is bounded by ${\cal B}(B^-\to
\bar {D}^{**0}K^-\to D^-\pi^+K^-) <5.0\times10^{-8}$, then for a
measured $CP$ asymmetry $|A_{CP}|=0.20$, we have the limit
$|\sin\gamma|>0.75$. For larger $CP$ asymmetries, the bound becomes
tighter.  The second solution
results only in larger values of $|\sin\gamma|$, consistent with the
bound obtained from the first solution. Once, the branching ratio,
${\cal B}(B^-\to \bar {D}^{**0}K^-\to D^-\pi^+K^-)$ and the direct $CP$
asymmetry are both measured, the value of $|\sin\gamma|$ corresponding to
each of the solutions can be extracted.

In the above estimates, we used the branching ratio for the decay of
$D^0$ to the particular $CP$ eigenstate $K_S\pi^0$. To improve
statistics we could add either all $CP$ even or all $CP$ odd states. In
fact, since the solution for $|\sin\gamma|$, involves only squares of
terms that switch sign from $CP$ even to $CP$ odd, it is possible to
achieve even higher statistics (increase $r^2$), by combining all
possible $CP$ eigenstate modes. Also, $D^+$ ($D^-$) could be
reconstructed using additional final states, which could increase
$a_c^2$ ($a_u^2$).

To conclude, we have suggested the use of $B\to D^{**0}K$ modes for
determination of $\gamma$. The $D^{**0}$ ($\bar{D}^{**0}$) decay to
flavour specific modes $D^+/D^{*+}\pi^-$ ($D^-/D^{*-}\pi^+$), allowing
measurement of the $b\to u$ amplitude. This overcomes the problem of
the Gronau-London-Wyler proposal and a clean extraction of $\gamma$ is
possible. The Belle collaboration has already measured the product of
branching fractions: ${\cal B}(B^-\to D^{**0}\pi^-) \times {\cal
B}(D^{**0}\to (D/D^*)^+\pi^-) \approx O(10^{-4})$, for four of the
D-meson excited states. A measurement of the branching ratios for the
$D^{**0}K^-$ final state should be feasible very soon. The large
number of $B-\bar{B}$ pairs expected at the high luminosity super-B
factories and other planned B-physics experiments, should enable a
measurement of the branching fractions for ${\cal B}(B^-\to
\bar{D}^{**0}K^-) \times {\cal B}(\bar{D}^{**0}\to (D/D^*)^-\pi^+)$,
allowing $\gamma$ to be cleanly extracted.  Even an upper limit on the
${\cal B}(B^-\to\bar{D}^{**0}K^-)$ will place an assumption free,
lower bound on $\gamma$.

{\bf Acknowledgements}: The author thanks G.~Rajasekaran, A.~Soffer,
O.~Long and D.~London for their comments. This work was supported by a
project under the Department of Science and Technology, India.


\begin{thebibliography}{References:}
\bibitem{Harrison:1998yr}
For a review, see for example, P.~F.~.~Harrison and H.~R.~.~Quinn (Eds.),
The BaBar Physics Book, SLAC Report 504, October 1998.
\bibitem{Browder:2003ii}
T.~E.~Browder, to appear in the proceedings of 21st International
Symposium on Lepton and Photon Interactions at High Energies (LP 03),
Batavia, Illinois, August 11-16, 2003 (http://conferences.fnal.gov/lp2003/program/papers/browder.pdf). 
arXiv:hep-ex/0312024. 
\bibitem{Gronau:2003br}
M.~Gronau and J.~L.~Rosner,
arXiv:hep-ph/0311280.
\bibitem{GLW}
M.~Gronau and D.~London.,
Phys.\ Lett.\ B {\bf 253}, 483 (1991);
M.~Gronau and D.~Wyler,
Phys.\ Lett.\ B {\bf 265} (1991) 172.
\bibitem{Atwood:1996ci}
D.~Atwood, I.~Dunietz and A.~Soni,
Phys.\ Rev.\ Lett.\  {\bf 78}, 3257 (1997)
[arXiv:hep-ph/9612433];
\bibitem{Sinha:1997zu} 
N.~Sinha and R.~Sinha,
Phys.\ Rev.\ Lett.\  {\bf 80}, 3706 (1998)
[arXiv:hep-ph/9712502].
\bibitem{dk}I.~Dunietz,
Phys.\ Lett.\ B {\bf 270}, 75 (1991); D.~Atwood, G.~Eilam, M.~Gronau and A.~Soni,
Phys.\ Lett.\ B {\bf 341}, 372 (1995); J.~H.~Jang and P.~Ko,
Phys.\ Rev.\ D {\bf 58}, 111302 (1998); M.~Gronau and J.~L.~Rosner,
Phys.\ Lett.\ B {\bf 439}, 171 (1998); M.~Gronau,
Phys.\ Rev.\ D {\bf 58}, 037301 (1998)
[arXiv:hep-ph/9802315];
 D.~Atwood and A.~Soni,
arXiv:hep-ph/0312100.
\bibitem{recent} Y.~Grossman, Z.~Ligeti and A.~Soffer,
Phys.\ Rev.\ D {\bf 67}, 071301 (2003)
[arXiv:hep-ph/0210433]; R.~Aleksan, T.~C.~Petersen and A.~Soffer,
Phys.\ Rev.\ D {\bf 67}, 096002 (2003)
[arXiv:hep-ph/0209194];
A.~Giri, Y.~Grossman, A.~Soffer and J.~Zupan,
Phys.\ Rev.\ D {\bf 68}, 054018 (2003)
[arXiv:hep-ph/0303187].
\bibitem{soffer}Talk given by A.~Soffer at Super B Factory Workshop
in Hawaii, Honolulu, January 19-22, 2004
(http://www.phys.hawaii.edu/~superb04/talks/Soffer.ppt).
\bibitem{expt} Ref.\cite{belle} and references therein. 
\bibitem{belle} K.~Abe {\it et al.}  [Belle Collaboration],
arXiv:hep-ex/0307021.
\bibitem{Wolfenstein:1983yz}
L.~Wolfenstein,
Phys.\ Rev.\ Lett.\  {\bf 51}, 1945 (1983).
\bibitem{Kayser:1999bu}
B.~Kayser and D.~London,
Phys.\ Rev.\ D {\bf 61}, 116013 (2000)
[arXiv:hep-ph/9909561].
\bibitem{Abi}A.~ Soffer, private communication.
\bibitem{numbers}The product ${\cal B}(B^-\to D_2^{*0}\pi^-) {\cal
      B}(D_2^{*0}\to D^+\pi^-)$ has been measured by the Belle
      Collaboration and is given in \cite {belle}. For all other
      branching ratios, we use the values listed in \cite{Hagiwara:fs}.
\bibitem{Hagiwara:fs}
K.~Hagiwara {\it et al.}  [Particle Data Group Collaboration],
Phys.\ Rev.\ D {\bf 66}, 010001 (2002).
\end{thebibliography}
\end{document}